\documentclass{aa}
\def\e{et~al.\ }
\def\es{{\rm erg\,s^{-1}} }
\def\r{R_{\rm m}}
\newcommand{\be}{\begin{equation}}
\newcommand{\ee}{\end{equation}}
\newcommand{\bdm}{\begin{displaymath}}
\newcommand{\edm}{\end{displaymath}}

\begin{document}

   \title{On the state of low luminous accreting neutron stars}

   \author{N.R. Ikhsanov\inst{1,2}}

 \offprints{N.R.~Ikhsanov\\ \email{ikhsanov@mpifr-bonn.mpg.de}}

   \institute{Max-Planck-Institut f\"ur Radioastronomie, Auf dem H\"ugel 69,
              D-53121 Bonn, Germany
              \and
              Central Astronomical Observatory of the Russian Academy of
              Science at Pulkovo,
              Pulkovo 65--1, 196140 Saint-Petersburg, Russia}

   \date{Received 5 March 2001/ Accepted 14 June 2001}

%\authorrunning{}
%\titlerunning{}

    \abstract{
Observational appearance of a neutron star in the {\it subsonic
propeller} state which is a companion of a wind-fed mass-exchange
close binary system is discussed. During the subsonic propeller state
(which was first introduced by Davies \e \cite{dfp79}) the neutron
star magnetosphere is surrounded by a spherical quasi-static
plasma envelope, which is extended from the magnetospheric boundary up
to the star accretion radius. The energy input to the envelope due to
the propeller action by the neutron star exceeds the radiative losses
and the plasma temperature in the envelope is of the order of the
free-fall temperature. Under this condition the magnetospheric
boundary is interchange stable. Nevertheless, I find that the rate of
plasma penetration from the envelope into the magnetic field of the
neutron star due to diffusion and magnetic field line reconnection
processes is large enough for the accretion power to dominate the
spindown power. I show that the accretion luminosity of the neutron
star in the subsonic propeller state is $L_{\rm a} \simeq 5 \times
10^{30} \div 10^{33} M_{15}\,{\rm erg\,s^{-1}}$, where $\dot{M}_{15}$
is the strength of the normal companion stellar wind which is
parametrized in terms of the maximum possible mass accretion rate onto
the neutron star magnetosphere. On this basis I suggest that neutron
stars in the subsonic propeller state are expected to be observed as
low luminous accretion-powered pulsars. The magnetospheric radius of
the neutron star in this state is determined by the strength of the
stellar wind, $\dot{M}_{\rm c}$, while the accretion luminosity is
determined by the rate of plasma penetration into the star
magnetosphere, $\dot{M}_{\rm a}$, which is $\dot{M}_{\rm a} \ll
\dot{M}_{\rm c}$. That is why the classification of the neutron star
state in these objects using the steady accretion model (i.e. setting
$\dot{M}_{\rm a} = \dot{M}_{\rm c}$) can lead to a mistaken
conclusion.
 }

   \maketitle

%%%%%%%%%%%%%%%%%%%%%%%%%%%%%%%%%%%%%%%%%%%%%%%%%%%%%%%%%%%%%

   \section{Introduction}

As was first recognized by Shvartsman (\cite{sh70}), three states
of a rotating, magnetized neutron star in a close binary system can
be distinguished: {\it ejector}, {\it propeller} and {\it accretor}.
This classification reflects three different evolutionary stages and
three different mechanisms of energy release responsible for the
neutron star emission. The sequence of states which a magnetized
neutron star in a wind-fed mass-exchange binary system follows as it
spins down from the initially very short periods can be expressed in
the form of the following chain: {\it ejector} $\rightarrow$ {\it
propeller} $\rightarrow$ {\it accretor}.

Neutron stars in the state of {\it ejector} are known as
spin-powered pulsars, in which the spindown power,
    \be\label{ll}
L_{\rm md}=\frac{2\,\mu^{2}\,\sin^{2}{\beta}}{3\,c^{3}}\,\omega^{4},
           \ee
dominates the star energy budget and is spent predominantly on the
generation of magneto-dipole waves and particle acceleration. Here
$\mu$ is the neutron star magnetic moment, $\beta$ is the angle
between the rotational and magnetic axes, $\omega=2 \pi/P_{\rm s}$ is
the star angular velocity, $P_{\rm s}$ is the neutron star spin
period and $c$ is the speed of light.

The state of a neutron star is classified as {\it accretor}
if the star luminosity is dominated by the accretion power,
    \be\label{la}
L_{\rm a} = \dot{M}_{\rm a} \frac{GM_{\rm ns}}{R_{\rm ns}}.
   \ee
Here $M_{\rm ns}$ and $R_{\rm ns}$ are the mass and the radius of
the neutron star, respectively, and $\dot{M}_{\rm a}$ is the mass
accretion rate onto the neutron star surface.

The intermediate state is called {\it propeller}. The necessity to
introduce this state can be justified as follows. The pulsar-like
spindown ceases when the pressure of the material being ejected
by the neutron star,
     \be\label{pout}
p_{\rm out} = \frac{L_{\rm md}}{4 \pi c R_{\alpha}^2},
    \ee
can no longer balance the pressure of the surrounding gas,
    \be\label{pinfty}
p_{\infty} = \frac{1}{2} \rho_{\infty} V_{\rm rel}^2.
    \ee
Here $\rho_{\infty}$ is the density of the surrounding gas,  $V_{\rm
rel}$ is the relative velocity between the neutron star and the
surrounding gas, and $R_{\alpha}$ is the accretion radius of the
neutron star,
    \be
R_{\alpha} = \frac{2 GM_{\rm ns}}{V_{\rm rel}^2}.
  \ee
Combining Eqs.~(\ref{pout}) and (\ref{pinfty}) one finds that this
occurs when the neutron star spin period $P_{\rm s} = P_{\rm md}$,
where
   \be\label{pmd}
P_{\rm md} = 0.3\ \mu_{30}^{1/2} V_8^{-1/4}\
(\sin{\beta})^{1/2}
\left[\frac{\dot{M}_{\rm c}}{10^{15}\,{\rm g\,s^{-1}}}\right]^{-1/4} \
{\rm s}.
     \ee
Here $\mu_{30}=\mu/10^{30}\,{\rm G\,cm^3}$, $V_8=V_{\rm
rel}/10^8\,{\rm cm\,s^{-1}}$ and $\dot{M}_{\rm c}$ is the strength
of the normal companion stellar wind which is parametrized in terms
of the maximum possible accretion rate following\footnote{There
should be no confusion between $\dot{M}_{\rm c}$, which is the mass
of the surrounding material interacting with the neutron star moving
through the wind in a time unit and $\dot{M}_{\rm a}$,
which is the mass accretion rate onto its surface. In the general
case $\dot{M}_{\rm c} \neq \dot{M}_{\rm a}$. The case of $\dot{M}_{\rm
c}=\dot{M}_{\rm a}$ corresponds to the steady accretion process,
when all material captured by the neutron star reaches the star
surface. In this particular case $\dot{M}_{\rm c}$ is called the rate
of mass capture by the neutron star (Bondi \& Hoyle \cite{bh44}).
However, as it will be shown below, it has a different physical
meaning if the neutron star is in the state of {\it propeller}} 
Davies \e (\cite{dfp79})
    \be\label{mc}
\dot{M}_{\rm c} = \pi R_{\alpha}^2 \rho_{\infty} V_{\rm rel}.
   \ee

On the other hand, for the steady accretion onto the neutron star
surface to realize the canonical magnetospheric radius, which is
defined by equating the ram pressure of the inflowing material with
the magnetic pressure due to the dipole field of the neutron star,
    \be\label{rm}
R_{\rm m} \equiv \left(\frac{\mu^{2}}{\dot{M}_{\rm c} \sqrt{2GM_{\rm
ns}}}\right)^{2/7},
     \ee
should be smaller than the corotational radius\footnote{This is
a necessary but not a sufficient condition},
      \be\label{rcor}
R_{\rm cor}= \left(\frac{GM_{\rm ns}}{\omega^2}\right)^{1/3}.
      \ee
Otherwise the neutron star proves to be in the centrifugal
inhibition regime (i.e. the  centrifugal acceleration at the
magnetospheric boundary, $\omega^2 \r$, dominates the gravitational
acceleration, $GM_{\rm ns}/\r^2$) and hence, the plasma accretion
onto the star surface is impossible. Putting
Eq.~(\ref{pmd}) to (\ref{rcor}) one finds that the condition $r_{\rm
m} \la R_{\rm cor}$ is satisfied only if $\dot{M}_{\rm c} \ga
\dot{M}_{\rm ea}$, where
     \be\label{mea}
\dot{M}_{\rm ea} \simeq 0.75\ \mu_{30}^2 V_8^{7/5} m^4
(\sin{\beta})^{-14/5}\ {\rm M_{\sun}\,yr^{-1}}.
    \ee
Here $m$ is the mass of the neutron star expressed in units of
$M_{\sun}$. However so huge a mass transfer rate can never be realized
in a wind-fed mass-exchange close binary system ($\dot{M}_{\rm ea}$
exceeds the maximum possible rate of mass loss by O-type supergiants
by more than five orders of  magnitude\,!). That is why the direct
{\it ejector} $\rightarrow$ {\it accretor} state transition under the
conditions of interest cannot be realized and the evolutionary track
of the initially fast rotating magnetized neutron star must contain
an additional spindown state which is traditionally called {\it
propeller}.

A neutron star in the state of propeller is spinning down due to the
interaction between its fast rotating magnetosphere and the
surrounding material. The mechanism of this interaction has been
investigated in detail by Davies \e (\cite{dfp79}) and Davies \&
Pringle (\cite{dp81}) who have shown that during the propeller state
the star magnetosphere is surrounded by a spherical quasi-static
envelope in which the plasma temperature is of the order of the
free-fall temperature,
    \be\label{tff}
T(R) \simeq T_{\rm ff}(R)=\frac{GM_{\rm ns} m_{\rm p}}{k R}.
   \ee
Here $m_{\rm p}$ is the proton mass and $k$ is the Boltzmann constant.
The rotational energy loss by the neutron star is convected up
through the envelope by the turbulent motions and lost through its
outer boundary. The structure of the envelope and the spindown rate
of the star depend on the value of the ratio:
   \bdm
\kappa=\frac{\omega\r}{V_{\rm s}(\r)},
   \edm
where $\r$ is the magnetospheric radius of the neutron star and
$V_{\rm s}(\r)$ is the sound speed in the envelope plasma, which
according to Eq.~(\ref{tff}) is of the order of the free-fall
velocity, $V_{\rm ff}$:
     \be\label{vs}
V_{\rm s}(\r) \simeq V_{\rm ff}(\r)=\sqrt{\frac{2GM_{\rm ns}}{\r}}.
    \ee
On this basis, Davies \e (\cite{dfp79}) distinguished three
sub-states of the propeller state\footnote{According to their
classification, case `a' corresponds to the pulsar-like
spindown, i.e. the ejector state.}: (b)~very rapid rotator ($\kappa
\gg 1$); (c)~supersonic propeller ($\kappa \ga 1$) and (d)~subsonic
propeller ($\kappa < 1$).

In cases `b' and `c' the magnetospheric radius of the neutron
star exceeds its corotational radius. The neutron star is in the
centrifugal inhibition regime and no accretion onto its surface
occurs. The spindown power dominates the energy budget and
is comparable to the radiative losses of the envelope.

When the neutron star spin period decreases below the critical
period,
    \be\label{pcd}
P_{\rm cd} = 23\ \mu_{30}^{6/7} m^{-5/7}
\left[\frac{\dot{M}_{\rm c}}{10^{15}\,{\rm g\,s^{-1}}}\right]^{-3/7}\
{\rm s},
   \ee
the corotational radius reaches the magnetospheric radius. Under the
condition $P_{\rm s} > P_{\rm cd}$ the wind plasma, 
penetrating from the base of the envelope into the star magnetic
field, is able to flow along the magnetic field lines and to
accrete onto the star surface. However the rate of plasma penetration
into the star magnetic field remains essentially smaller than 
$\dot{M}_{\rm c}$ if the magnetospheric boundary is stable with
respect to interchange instabilities (e.g. Rayleigh-Taylor
instability). According to Arons \& Lea (\cite{al76a}) and  Elsner \&
Lamb (\cite{el76}), the onset condition for the interchange
instability of the magnetospheric boundary reads
    \be\label{tcr}
 T(\r) < T_{\rm cr} \simeq 0.3\ T_{\rm ff}.
    \ee
This means that the steady accretion process under the condition
$P_{\rm s} > P_{\rm cd}$ can be realized only if the cooling of the
envelope plasma (due to the radiation and convective motions)
dominates the energy input to the envelope due to the propeller
action by the neutron star.

Investigating this particular state, Davies \& Pringle  (\cite{dp81},
hereafter DP81) have shown that the cooling processes in the envelope
are effective if the spin period of the star exceeds a so-called
break period $P_{\rm br}$, which according to Ikhsanov (\cite{i01a})
is
   \be\label{pbr}
P_{\rm br} \simeq\ 450\ \mu_{30}^{16/21}\ \dot{M}_{15}^{-5/7}\
m^{-4/21}\ {\rm s}.
      \ee
DP81 have assumed that during the epoch when $P_{\rm
cd} < P_{\rm s} < P_{\rm br}$, no accretion onto the star surface is
realized (i.e. $\dot{M}_{\rm a}=0$) and the star energy budget is
dominated by the spindown power, which in the case of this -- {\it
subsonic propeller} -- state is
   \be\label{lssp}
L_{\rm ssp} = 6.4 \times 10^{31}\,{\rm egr\,s^{-1}}\ \mu_{30}^2\
m^{-1} P_{25}^{-3},
   \ee
where $P_{25}=P_{\rm s}/25$\,s.

At the same time, the assumption $\dot{M}_{\rm a}=0$ under the
condition $\r < R_{\rm cor}$ is not obvious since the interchange
instability is not the only mode by which the accreting plasma can
enter the neutron star magnetic field. In this paper I show that
the accretion luminosity of the neutron star in the subsonic
propeller state can be comparable or even exceeds the
spindown power if the  plasma penetration into the magnetosphere is
governed by the magnetic field lines reconnection process (section~2).
In this case the star can be observed as a low luminous
accretion-powered X-ray pulsar rather than the spin-powered propeller.
The results are discussed and summarized in section~3.

    \section{Accretion luminosity of a neutron star in `subsonic
propeller' state}

I consider a close binary system in which a rotating, magnetized
neutron star is orbiting around a high-mass main sequence companion
which underfills its Roche lobe and loses mass in the form of
stellar wind. The neutron star, which moves through the
wind of the normal companion, is assumed to be in the state of {\it
subsonic propeller}. This means that the neutron star magnetosphere
is surrounded by a hot plasma envelope ($T\approx T_{\rm ff}$) which
is extended from the magnetospheric boundary, $\r \la R_{\rm
cor}$, up to the accretion radius of the neutron star, $R_{\alpha}$.
The spin period of the neutron star lies within the interval $P_{\rm
cd} < P_{\rm s} < P_{\rm br}$, and thus, the rate of energy input
to the envelope due to the propeller action by the neutron star
(expressed by Eq.~\ref{lssp}) dominates the radiative losses from the
envelope. Under this condition the envelope can be considered as a
static adiabatic atmosphere in which the plasma pressure $p \propto
R^{-5/2}$, the sound speed $c_{\rm s} \propto R^{-1/2}$ and the
number density $n \propto R^{-3/2}$. The plasma pressure at the outer
edge of the envelope is equal to the ram pressure of the surrounding
gas. The formation of the envelope in the first approximation prevents
the stellar wind from penetrating into the accretion lobe of the
neutron star (i.e. the sphere with the radius $R_{\alpha}$).
In this case stellar wind overflows the outer edge with the rate
$\dot{M}_{\rm c}$
compressing and heating the envelope plasma (for discussion see DP81).

The equilibrium shape of a neutron star magnetosphere surrounded by
a hot hydrogen envelope has been reconstructed by Arons \& Lea
(\cite{al76a}), Elsner \&Lamb (\cite{el77}) and Michel (\cite{m77a}).
All these authors have argued that the magnetosphere in the
considered case tends to be closed and its boundary is convex
towards the accreting plasma, with two cusp points situated on the
magnetic axis of the neutron star. It is rotationally symmetric about
the  magnetic axis and reflection symmetric about the equatorial
plane.

      \subsection{The rate of plasma entry into the star
magnetosphere}

Formation of the magnetosphere, to a first approximation,
prevents the envelope plasma from reaching the neutron star itself.
That is why the rate of mass accretion onto the neutron star surface
strongly depends on the rate of plasma penetration into the
magnetosphere at its boundary. As was been mentioned in the 
introduction, under the condition $P_{\rm s} < P_{\rm br}$, the
cooling of the envelope plasma is not effective and hence the
magnetospheric boundary is interchange stable. Under this
condition the modes by which the envelope plasma can enter the star
magnetosphere are (i)~turbulent diffusion and (ii)~reconnection of
the magnetic field lines.

           \subsubsection{Diffusion}

The conductivity of the envelope plasma can be approximated by
that of a fully ionized gas: $\sigma \sim 10^7 T_{\rm
e}^{3/2}$, where $T_{\rm e}$ is the electron temperature. In our
case, the value of conductivity is very high but, nevertheless,
finite ($\sigma < \infty$). That is why a diffusion skin layer
(magnetopause) at the magnetospheric boundary  is formed. The
thickness of the magnetopause in the general case depends on the
velocity and the time of plasma diffusion across the magnetic field
lines, but in any case it is larger than the Larmor radius of ions.
Under the condition $P_{\rm cd} \ga P_{\rm s}$, the gravitation
force applied to the plasma in the diffusion layer dominates the
centrifugal force. Therefore the plasma penetrating the layer
flows along the field lines toward the magnetic poles of the
neutron star and accretes onto the star surface.

The rate of plasma flow into the magnetopause from the envelope
due to diffusion is
   \be\label{dotmin}
\dot{M}_{\rm d} \simeq 4 \pi \r^2\ \rho(\r)\ V_{\rm diff}.
   \ee
$\rho(\r)$ is the plasma density at the base of the envelope just
over the magnetosphere boundary. It can be evaluated equating the
plasma ram pressure and the magnetic field pressure at the boundary:
   \be
\rho(\r) = \frac{\mu^2}{4 \pi GM_{\rm ns} \r^5}.
  \ee
$V_{\rm diff}$ is the diffusion velocity which, according to
Ikhsanov \& Pustil'nik (\cite{ip96}), can be expressed in the
following form
    \be
 V_{\rm diff} = \sqrt{D_{\rm eff}/t_{\rm ff}}.
   \ee
Here $D_{\rm eff}$ is the effective diffusion coefficient and
$t_{\rm ff}=\r^{3/2}/\sqrt{GM_{\rm ns}}$ is the free-fall time, which
is the characteristic time of the plasma flow from the layer to the
star surface.

In the general case the value of $D_{\rm eff}$ is obviously limited
by the following relation:
    \bdm
D_{\rm cr} \la D_{\rm eff} \la D_{\rm B},
   \edm
where $D_{\rm cr} = c^2/4 \pi \sigma$ and $D_{\rm B} = \zeta V_{\rm
T(i)} r_{\rm h(i)}$ are the coefficients of the classical and Bohm
diffusion, respectively, $V_{\rm T(i)}$ and $r_{\rm h(i)}$ are the
thermal velocity and Larmor radius of ions, and $\zeta$ is the
diffusion efficiency.

The minimum rate of plasma entry into the magnetosphere can be
obtained putting $D_{\rm eff} \sim D_{\rm cl}$. In this case one finds
 \be
\dot{M}_{\rm cl} \simeq 10^6\ \mu_{30}^{2/7}\ m^{-15/14}
\left(\frac{\dot{M}_{\rm c}}{10^{15}\,{\rm g\,s^{-1}}}\right)^{6/7}\
{\rm g\,s^{-1}}.
 \ee

The maximum inflow rate is realized if the diffusion process is
governed by drift-dissipative instabilities (i.e. Bohm diffusion)
  \be\label{bohm}
D_{\rm eff} \simeq D_{\rm B} = \frac{\zeta c k T_{\rm i}(\r)}{16 e
B(\r)},
    \ee
where $T_{\rm i}$ is the ion plasma temperature at the base of the
envelope. Combining Eqs.~(\ref{dotmin} -- \ref{bohm}) yields
  \be\label{dotmdif}
\dot{M}_{\rm B} \sim 3\,10^{10}\,{\rm g\,s^{-1}}\
\zeta_{0.1}^{1/2}\ \mu_{30}^{-1/14} m^{1/7}
\left(\frac{\dot{M}_{\rm c}}{10^{15}\,{\rm g\,s^{-1}}}\right)^{11/14}
  \ee
where $\zeta_{0.1}=\zeta/0.1$. This normalization of the diffusion
efficiency is adopted following the results of experiments on the
nuclear fusion (e.g. Hamasaki \e \cite{ham74}) and the measurements
of the solar wind penetrating the magnetosphere of the Earth
(Gosling \e \cite{gosling91}).

Hence, the total rate of mass flow into the magnetosphere due to the
diffusion process is $\dot{M}_{\rm d}=\dot{M}_{\rm cl} + \dot{M}_{\rm
B}$.

            \subsubsection{Reconnections}

According to Elsner \& Lamb (\cite{el84}) the rate of plasma entry
into the magnetosphere due to reconnection of the magnetic field
lines can be expressed as
    \be
\dot{M}_{\rm rec} \simeq
\alpha_{\rm r} \frac{\mu^2}{GM_{\rm ns} \r^3}\
\frac{A_{\rm r}}{4 \pi \r^2} V_{\rm a},
  \ee
where $\alpha_{\rm r}$ is the efficiency of the reconnection process,
$A_{\rm r}= 4 \pi \r \lambda_{\rm m}$ is the effective area of the
reconnection region and $V_{\rm a}=B/\sqrt{4 \pi \rho}$ is the Alfv\'en
velocity, which in the boundary layer at the base of the hot envelope is
of the order of $V_{\rm ff}(\r)$.

The values of $\alpha_{\rm r}$ and $A_{\rm r}$ depend on the current
sheet parameters and the magnetic field configuration in the
accretion flow.
Investigations of reconnection processes in solar flares and in the
Earth's magnetopause suggest the average value of $\alpha_{\rm r}
\approx 0.1$ (see Priest \& Forbes \cite{pf00} and references
therein).

The effective area of the reconnection region depends of the scale of the
magnetic field in the accreting plasma over the magnetospheric boundary.
It is limited by the initial inhomogeneity of the accretion flow
and/or by the fragmentation of plasma over the rotating magnetospheric
boundary due to the Kelvin-Helmholtz instability.
In the latter case the average scale of plasma vortices of the
embedded field is $\lambda_{\rm m} \sim 0.1 \r$ (see Arons \& Lea
\cite{al76b}; Wang \& Robertson \cite{wr85}). On the other hand, the
scale of the plasma inhomogeneity at the base of the envelope can be
evaluated by the size of the convective cell (i.e. by the height of
the homogeneous atmosphere over the magnetospheric boundary, $h_{\rm
g}$). In our case ($T \ga 0.3 T_{\rm ff}$) one finds $h_{\rm g} \ga
0.3 \r$ and, correspondingly, the ratio of the area of a convective
cell to the total area of the magnetospheric boundary is
$(h_{\rm g}/\r)^2 \sim 0.1$. In this context the estimate
$\lambda_{\rm m} \sim 0.1 \r$ seems to be appropriate.

Under these conditions the rate of plasma penetration into the neutron
star magnetic field due to the magnetic lines reconnection reads
  \be\label{dotmrec}
\dot{M}_{\rm rec} \simeq 10^{13}\,{\rm g\,s^{-1}}\
\left[\frac{\alpha_{\rm r}}{0.1}\right]
\left[\frac{\lambda_{\rm m}}{0.1 \r}\right]
\left(\frac{\dot{M}_{\rm c}}{10^{15}\,{\rm g\,s^{-1}}}\right).
  \ee

Thus, the rate of plasma penetration into the magnetosphere of the
neutron star in the state of {\it subsonic propeller} can be expressed
as
   \be\label{madrec}
\dot{M}_{\rm a} = \dot{M}_{\rm d} + \dot{M}_{\rm rec}.
  \ee

       \subsection{Accretion luminosity}

Under the condition $\r < R_{\rm cor}$ plasma penetrating the
magnetic field of the neutron star is accreted onto its surface.
This indicates that the observed total luminosity of the neutron star
in the state of subsonic propeller is
    \be
L_{\rm bol} = L_{\rm ssp} + L_{\rm a}.
   \ee
The lower limit to the accretion luminosity, $L_{\rm a}$, can be
estimated assuming the plasma entry in the stellar magnetic field
from the base of the envelope due to diffusion (see
Eq.~\ref{dotmdif}):
       \be\label{lad}
L_{\rm a, diff } \sim 4\,10^{30}\ \zeta_{0.1}^{1/2}
\mu_{30}^{-1/14} m^{8/7} R_6^{-1} \dot{M}_{15}^{11/14}\
{\rm erg\,s^{-1}},
      \ee
where $\dot{M}_{15}$ is the strength of the stellar wind expressed in
units of $10^{15}\,{\rm g\,s^{-1}}$.

Combining Eqs.~(\ref{lssp}) and (\ref{lad}) one finds that the
accretion power $L_{\rm a, diff}$ dominates the star energy budget if
$P_{\rm s} \ga P_{\rm dda}$, where
     \be
P_{\rm dda} = 63\ \zeta_{0.1}^{1/6}\ \mu_{30}^{29/42} m^{-5/7} 
R_6^{1/3} \dot{M}_{15}^{-11/42}\ {\rm s}.
    \ee

If the plasma penetration the magnetosphere of a neutron star is
governed by the reconnection of the magnetic field lines, 
the accretion luminosity of the star is
     \be
L_{\rm a, rec} \simeq 10^{33}\ \alpha_{0.1}\ m\ R_6^{-1}\
\lambda_{0.1} \dot{M}_{15} {\rm erg\,s^{-1}},
  \ee
where $\alpha_{0.1}=\alpha_{\rm r}/0.1$ and
$\lambda_{0.1}=\lambda_{\rm m}/0.1 \r$. In this case the accretion
power already dominates the spindown power at $P_{\rm cd}$ if 
$\dot{M}_{\rm c} > \dot{M}_{\rm rda}$, where
   \be
\dot{M}_{\rm rda} \simeq 6\,10^{13}\ \mu_{30}^2\ m^{-2}\ R_6\
\alpha_{0.1}^{-1}\ \lambda_{0.1}^{-1}\ P_{25}^{-3}\,{\rm g\,s^{-1}}.
   \ee

In both cases the neutron star will be recognized as a low luminous
accretion-powered pulsar in which the radius and the temperature of
polar caps are\footnote{The black body radiation is assumed}
    \bdm
R_{\rm pc}= 0.3\ \mu_{30}^{-2/7} m^{1/14}
R_6^{3/2} \left[\frac{\dot{M}_{\rm c}}{10^{15}\,{\rm
g\,s^{-1}}}\right]^{1/7}\ {\rm km},
    \edm
    \bdm
T_{\rm pc} \simeq 8 \times 10^6\ \
\left[\frac{R_{\rm pc}}{0.3\,{\rm km}}\right]^{-1/2}
\left[\frac{L_{\rm a}}{10^{33}\,{\rm erg\,s^{-1}}}\right]^{1/4}\
{\rm K}.
    \edm

   \subsection{Quasi-static envelope approximation}

The model of the envelope which is surrounding the magnetosphere of
a neutron star in the subsonic propeller state has been constructed by
DP81 under the assumption that no accretion onto the star surface
occurs. In this case the envelope can be considered as quasi-static,
i.e. the average rate of mass transfer through the envelope in the
radial direction is assumed to be zero. Under this condition the
plasma of the stellar wind does not penetrate into the accretion lobe
of the neutron star but overflows the outer edge of the envelope with
the rate $\dot{M}_{\rm c}$, compressing the envelope plasma.

One faces however a slightly different situation taking into account
that the ``magnetic gates'' during the subsonic propeller state of the
neutron star are not closed completely. As has been shown above,
the rate of plasma inflow into the interchange stable magnetosphere
due to diffusion and the magnetic field line reconnection  
is $\dot{M}_{\rm a} \neq 0$. In this case the envelope
remains in an equilibrium state if the amount of material inflowing
from the base of the envelope to the magnetosphere is compensated by
the same amount of material coming into the envelope through its outer
boundary. This means that the radial drift of plasma through the
envelope with the rate $\dot{M}_{\rm a}$ is expected. Hence the
question about the structure of this modified envelope arises.

The characteristic time of the accretion process in the envelope can
be estimated as
    \be\label{tdr}
t_{\rm dr} = R/V_{\rm dr},
   \ee
where $V_{\rm dr}$ is the average velocity of plasma radial drift
through the envelope, which under the condition $\dot{M}_{\rm
a}=$\,const, can be evaluated as
    \be
V_{\rm dr} = \frac{\dot{M}_{\rm a}}{\dot{M}_{\rm c}} V_{\rm ff}.
    \ee
On the other hand, the time scale on which the incoming energy
is redistributed within the envelope (i.e. the time of
envelope relaxation to the equilibrium state) is determined by the
characteristic turbulence time
     \be\label{tt}
t_{\rm t} = R/V_{\rm t}.
   \ee
Hence the quasi-static approximation can be applied to the modified
model of the envelope if $t_{\rm t} \ll t_{\rm dr}$. Combining
Eqs.~(\ref{tdr}) and (\ref{tt}) I find this condition to be
satisfied if the spin period of the star is $P_{\rm s} \ll P_{\rm
qs}$, where
    \bdm
P_{\rm qs} \simeq 1.6\,10^3 \mu_{30}^{6/7}
m^{-5/7}  \left[\frac{\dot{M}_{\rm a}}{10^{13}\,{\rm
g\,s^{-1}}}\right]^{-1} \left[\frac{\dot{M}_{\rm c}}{10^{15}\,{\rm
g\,s^{-1}}}\right]^{4/7} {\rm s}.
    \edm
This indicates that under the conditions of interest the structure of
the modified envelope (with the radial plasma drift)  can be
considered within the quasi-static approximation and thus,
the model presented by DP81 can be applied.

         \section{Discussion}

The basic conclusion of the investigation presented in this paper is
that the observational appearance of a neutron star in the state of
{\it subsonic propeller} strongly depends on the mode by which the
envelope plasma enters the magnetosphere of the star. The canonical
notion that the star in this state should be observed as a
non-pulsating soft X-ray source of the luminosity $L_{\rm ssp}$
is valid only in the particular case of the classical (Coulomb)
diffusion. If, however, the penetration process is governed by the
Bohm diffusion or/and the reconnections of the magnetic field lines,
the mass accretion rate onto the star surface proves to be high
enough for the accretion power to dominate the energy budget of the
star. In this case the neutron star is expected to be observed as
a steady low luminous ($L_{\rm x}=L_{\rm a, diff}+L_{\rm a, rec}$)
accretion-powered pulsar in which most of the X-ray emission comes
from the hot ($T_{\rm cp}$) spots of the radius $\sim R_{\rm cp}$
situated in the magnetic pole regions on the surface of the neutron
star (see Sect.\,2.2).

The entry of the accreting plasma into the interchange stable
magnetosphere of a neutron star has been investigated by
Elsner \& Lamb (\cite{el84}). They have shown that there are no
grounds for assuming that the Bohm diffusion or the field line
reconnections modes of the plasma penetrating the magnetosphere
are suppressed. They evaluated the efficiency of the penetration
process due to these modes to be of about 1\% of the efficiency of
that due to the interchange instabilities. This is too small to
explain the observed luminosities of the brightest X-ray pulsars
($\sim 10^{37}\div 10^{38}\es$) but is quite sufficient for the
interpretation of the relatively low luminous X-ray sources
($L_{\rm x} < 10^{36}\es$). The efficiency of the plasma penetrating
the magnetosphere of a neutron star evaluated by Elsner \&
Lamb is comparable to that obtained in laboratory experiments
(e.g. Hamasaki \e \cite{ham74}) and derived from the observations
of the solar wind entering the magnetosphere of the Earth
(e.g. Gosling \e \cite{gosling91}). The same results were obtained
in the simulations of the solar wind penetration of the Earth's
magnetic field due to diffusion (Winske \& Omidi \cite{wo95})
and reconnection of the field lines (Usadi \e \cite{usadi93}).
Thus the values $\zeta=0.1$ and $\alpha=0.1$ used in this paper
seem quite reasonable. The estimation of the parameter $\lambda$ is
more complicated and requires detailed study of the
convective process in the envelope. This investigation, however,
is beyond the scope of the present paper. Here I would like only
to note that the estimate of the magnetic field scale by the size
of a convection element is frequently used in the simulations of
the magnetic field generation in astrophysical objects (e.g.
accretion disks, active regions of the Sun, etc.).

The last point I would like to address here is the determination of
the state of a low luminous neutron star by the ratio between its
magnetospheric and corotation radii, $\Upsilon = \r/R_{\rm cor}$.
According to Eqs.~(\ref{rcor}) and (\ref{rm}) the value of
$\Upsilon$ depends on the mass, the spin period and the dipole
magnetic moment of the neutron star, and on the strength of the
stellar wind in which the neutron star is situated, $\dot{M}_{\rm
c}$. The parameters of the neutron star can be evaluated if it is
a component of a binary system and is observed as a pulsar
with the cyclotron line feature in the X-ray spectrum. In this case
the value of $\Upsilon$ can be obtained combining Eqs.~(\ref{la}),
(\ref{rm}) and (\ref{rcor}):
     \be\label{ypsilon}
\Upsilon = \frac{3 \xi^{2/7} \mu^{4/7}}{L_{\rm a}^{2/7} R_{\rm
ns}^{2/7} (GM_{\rm ns})^{4/21} P_{\rm s}^{2/3}}.
   \ee
Here, $\xi = \dot{M}_{\rm a}/\dot{M}_{\rm c}$ is the efficiency of
the accreting plasma penetrating the magnetic field of the star at
the magnetospheric boundary.

As a rule, the value of $\xi$ is assumed to be equal to unity.
This assumption is rather well justified if the neutron star
undergoes disk accretion or/and if its X-ray luminosity
exceeds the critical value,
     \bdm
L_{\rm cr} = 3\,10^{36}\ \mu_{30}^{1/4}\ m^{1/2}\ R_6^{-1/8}\ \es,
   \edm
at which the cooling of the plasma over the magnetospheric
boundary is effective and the boundary is interchange unstable
(see Ghosh \& Lamb \cite{gl79}; Arons \& Lea \cite{al76a}).

If, however, the plasma penetration into the magnetosphere of
a {\it spherically} accreting neutron star is governed by the
diffusion or/and the field lines reconnection processes, the
assumption $\xi \simeq 1$ is not valid. In particular, in the case
of subsonic propeller $\xi \la 0.01$ and hence the value of
$\Upsilon$ proves to be at least a factor of four smaller than
that evaluated using the traditional method:
  \bdm
\Upsilon(\xi) = 0.27  \left(\frac{\xi}{0.01}\right)^{2/7}
\Upsilon(\xi=1).
    \edm
This indicates that the overestimation of the parameter $\xi$ may
lead to the mistaken classification of an accretion-powered source
as supersonic propeller. An example of such a situation in the
analysis of the low luminous state of the neutron star in the
Be/X-ray transient A0535+26 has been recently presented by the
author (Ikhsanov \cite{i01b}).

 \begin{acknowledgements}
I would like to thank the anonymous referee for suggested
improvements. I acknowledge the support of the Follow-up  program of
the Alexander von Humboldt Foundation.
\end{acknowledgements}


\begin{thebibliography}{}
\bibitem[1976a]{al76a}
  Arons J., Lea S.M., 1976a, ApJ 207, 914
\bibitem[1976b]{al76b}
  Arons J., Lea S.M., 1976b, ApJ 210, 792
\bibitem[1944]{bh44}
  Bondi H., Hoyle F., 1944, MNRAS 104, 273
\bibitem[1979]{dfp79}
  Davies R.E., Fabian A.C., Pringle J.E., 1979, MNRAS 186, 779
\bibitem[1981]{dp81}
  Davies R.E., Pringle J.E., 1981, MNRAS 196, 209
\bibitem[1976]{el76}
  Elsner R.F., Lamb F.K., 1976, Nature 262, 356
\bibitem[1977]{el77}
  Elsner R.F., Lamb F.K., 1977, ApJ 215, 897
\bibitem[1984]{el84}
  Elsner R.F., Lamb F.K., 1984, ApJ 278, 326
\bibitem[1979]{gl79}
  Ghosh P., Lamb F.K., 1979, ApJ 232, 259
\bibitem[1991]{gosling91}
  Gosling J.T., Thomsen M.F., Bame S.J., \e, 1991, J. Geophys. Res.
96, 14097
\bibitem[1974]{ham74}
  Hamasaki S., Davidson R.C., Krall N.A. and Liewer P.C., 1974, Nucl.
  Fusion 14, 27
\bibitem[2001a]{i01a}
  Ikhsanov N.R., 2001a, A\&A 368, L5
\bibitem[2001b]{i01b}
  Ikhsanov N.R., 2001b, A\&A 367, 549
\bibitem[1996]{ip96}
  Ikhsanov N.R., Pustil'nik L.A., 1996, A\&A 312, 338
\bibitem[1977]{m77a}
  Michel F.C., 1977, ApJ 213, 836
\bibitem[2000]{pf00}
   Priest, E.R., Forbes T.G., 2000, ``Magnetic reconnection: MHD
  theory and applications'', Cambridge University Press
\bibitem[1970]{sh70}
  Shvartsman V.F., 1970, Radiofizika 13, 1852
\bibitem[1993]{usadi93}
  Usadi A., Kageyama A., Watanabe K., and Sato T., 1993, J.
Geophys. Res. 98, 7503
\bibitem[1985]{wr85}
    Wang Y.-M., Robertson J.A., 1985, A\&A 151, 361
\bibitem[1995]{wo95}
   Winske D., Omidi N., 1995, J. Geophys. Res. 100, 11923
\end{thebibliography}
\end{document}